\documentclass[aps,prl,superscriptaddress,preprintnumbers,%
               twocolumn,groupedaddress]{revtex4}
\usepackage{graphicx}
\usepackage{amsmath}
\usepackage{bm}
\begin{document}

\voffset=.25in
\preprint{\vbox{ \hbox{hep-ph/0305269}
                \hbox{ANL-HEP-PR-03-024}
                 \hbox{JLAB-THY-03-33}
                 \hbox{MIT-CTP 3380}
                 \hbox{SLAC-PUB-9861}
                 \hbox{YITP-SB-03-23}
         }}
\title{
Hunting for glueballs in electron-positron annihilation
}

\affiliation{High Energy Physics Division, Argonne National Laboratory,
Argonne, Illinois 60439}
\affiliation{Theory Group, Thomas Jefferson National Accelerator Facility, 
Newport News, Virginia 23036}
\affiliation{Center for Theoretical Physics,
Massachusetts Institute of Technology,
Cambridge, Massachusetts 02139}
\affiliation{Stanford Linear Accelerator Center, Stanford
University, Stanford, California 94309}
\affiliation{C.~N.~Yang Institute for Theoretical Physics,
State University of New York, Stony Brook, New York 11794-3840}

\author{Stanley J. Brodsky}
\affiliation{Theory Group, Thomas Jefferson National Accelerator Facility, 
Newport News, Virginia 23036}
\affiliation{Stanford Linear Accelerator Center, Stanford
University, Stanford, California 94309} 

\author{Alfred Scharff Goldhaber}
\affiliation{Center for Theoretical Physics,
Massachusetts Institute of Technology,
Cambridge, Massachusetts 02139}
\affiliation{Stanford Linear Accelerator Center,
Stanford University, Stanford, California 94309}
\affiliation{C.~N.~Yang Institute for Theoretical Physics,
State University of New York, Stony Brook, New York 11794-3840}

\author{Jungil Lee}
\affiliation{High Energy Physics Division, Argonne National Laboratory,
Argonne, Illinois 60439}
\affiliation{Theory Group, Thomas Jefferson National Accelerator Facility, 
Newport News, Virginia 23036}


\date{\today}
\begin{abstract}

We calculate the cross section for the exclusive production of
$J^{PC}=0^{++}$ glueballs 
$\mathcal{G}_0$ 
in association with the $J/\psi$ in $e^+e^-$ annihilation using the pQCD factorization formalism.  The required
long-distance matrix element for the glueball is bounded by  
CUSB data from a search for resonances in  radiative 
$\Upsilon$ decay.  The cross section for $e^+e^-\to
J/\psi+\mathcal{G}_0$  at 
$\sqrt{s}=10.6$~GeV is similar to exclusive  charmonium-pair
production 
$e^+e^-\to J/\psi+h$ for
$h=\eta_c$ and
$\chi_{c0}$, and is larger by a factor 2 than that for
$h=\eta_{c}(2S)$.  As the  subprocesses
$\gamma^* \to (c\bar c) (c \bar c)$  and 
$\gamma^* \to (c \bar c) (g g)$ are of the same nominal
order in perturbative QCD, it is possible that
some portion of the anomalously large signal observed by Belle in 
$e^+ e^- \to
J/\psi X$ may actually be due to the production of charmonium-glueball 
$J/\psi \mathcal{G}_J$ pairs. 
\end{abstract}



\maketitle


Bound states of gluons provide an explicit signature of the
non-Abelian interactions of quantum chromodynamics.  In fact, in a
model universe without quarks, the hadronic spectrum of QCD would
consist solely of color-singlet glueball states.  In the physical
world, the purely gluonic components mix with $q \bar q$ pairs,
leading to an enriched spectrum of isospin-zero states as well
as $q \bar q g$ hybrids.  The existence of this exotic spectrum is
as essential a prediction of QCD as the Higgs particle is for the
electroweak theory.

Lattice gauge theory predicts the spectrum and quantum numbers of
gluonic states.  
According to a recent 
calculation~\cite{Morningstar:1999rf}, 
the ground-state masses
for the $J^{PC}=0^{++}$ and $2^{++}$  glueballs $\mathcal{G}_J$ are
$1.73$ and $2.40$~GeV, respectively.
Thus far, the empirical
evidence for glueballs is not decisive, probably because of
complications from mixing with the quark degrees of freedom, but there
are indications of an extra neutral scalar state perhaps due to a
glueball of mass (before mixing)
near 1.7 GeV
\cite{close}. 

\begin{figure}
\includegraphics[height=3.5cm]{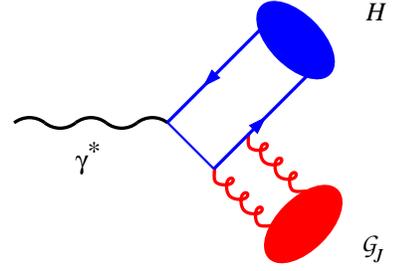}
\caption{\label{fig1}%
Feynman diagram for
$\gamma^* \to H + \mathcal{G}_J$.  }
\end{figure}

An important mechanism for producing glueballs is the radiative
decay of heavy quarkonium, particularly $J/\psi \to \gamma \mathcal{G}_J$ and
$\Upsilon \to \gamma \mathcal{G}_J$~\cite{Brodsky:1977du}.  In these
reactions, the quarkonium decays to an intermediate $\gamma g g$
state which then can couple to any charge conjugation parity $C=+$ 
isospin $I=0$ gluonic or hybrid
state.  
For example, the BES Collaboration~\cite{Shen:2002nv} has
observed the radiative decays of the $J/\psi$ and the $\psi(2S)$ to
$\gamma f_0(1710)$, a
glueball candidate.  
In this Letter we focus on another optimal mechanism for the
production of $\mathcal{G}_0$ and $\mathcal{G}_2$ at $e^+ e^-$ colliders,  
the reaction 
$e^+ e^- \to \gamma^* \to H \mathcal{G}_J$, $H=J/\psi$ or $\Upsilon$
\cite{gogo},  in which a $C=+$ glueball can be produced in association
with a quarkonium state from the subprocess 
$\gamma^* \to (Q \overline{Q}) (gg)$.  Two-gluon
components in $\eta$ particles have been estimated recently
\cite{hyb}.   One of the six Feynman diagrams for the subprocess is
shown in Fig.~\ref{fig1}; the remaining diagrams are permutations of the
photon and the two gluons.  A related reaction 
$\gamma^* \to \pi^0 \mathcal{G}_J$
has been considered~\cite{Wakely:1991ej}
as a source of
pseudoscalar glueballs.  We shall show that these reactions satisfy 
perturbative QCD~(pQCD)
factorization.
Unlike radiative quarkonium decay, this channel imposes no {\it a priori}
limit on the mass of the glueball.

The main background to charmonium-glueball production 
$e^+ e^- \to J/\psi \mathcal{G}_J$  is
exclusive quarkonium pairs such as $\gamma^* \to J/\psi \eta_c,$ arising
from the subprocess $\gamma^* \to (c\bar c) (c \bar c).$
The exclusive production of charmonium pairs has in fact
been observed recently with a substantial rate at
Belle~\cite{Abe:2002rb}.  The
rates for exclusive charmonium-pair production 
reported by Belle are significantly
larger than  predictions based on pQCD~\cite{Braaten:2002fi,Liu:2002wq}.
The Belle experiment identifies one member of the pair, the $J/\psi,$ via its
leptonic decay;
the other quarkonium state is inferred by identifying the missing mass of
the spectator system with the
charmonium states $\eta_c$, $\chi_{c0}$, and $\eta_c(2S)$ which occur within
the detector mass resolution.
As noted in Refs.~\cite{Bodwin:2002fk,Bodwin:2002kk}, 
some of the Belle signal for quarkonium pairs
may be due to two-photon annihilation 
$e^+ e^- \to \gamma^* \gamma^* \to J/\psi J/\psi$.  
Here we note that because the  subprocesses
$\gamma^* \to (c\bar c) (c \bar c)$  and 
$\gamma^* \to (c \bar c) (g g)$ are of the same nominal
order in pQCD,
it is possible that
some portion of the signal observed by Belle in $e^+ e^- \to J/\psi X$
may actually be due to the production of 
$J/\psi \mathcal{G}_J$ pairs.

In general, exclusive amplitudes can be computed in QCD by convoluting
the light-front wavefunctions 
$\psi_{n/H}(x_i,\textbf{k}_{\perp i})$ of each hadron
with the corresponding
$n$-particle irreducible quark-gluon matrix elements, summed over
$n$~\cite{Lepage:1980fj}.  For hadronic amplitudes involving a hard
momentum transfer
$Q$, it is usually possible to expand the quark-gluon scattering
amplitude as a function of ${\textbf{k}^2_\perp}/{Q^2}$.  The leading-twist
contribution can then be computed from a hard-scattering amplitude
$T_H$ where the external quarks and gluons associated with each
hadron are collinear.
Furthermore, only the minimum number of quark and
gluon quanta contribute at leading order in $1/Q^2.$
In our case, the relevant hard-scattering amplitude is
$T_H(\gamma^* \to c \bar c g g)$ computed with collinear $c$ and $\bar c$ and
collinear $g g.$
As $T_H$ at leading twist is independent of the constituent's relative
transverse momentum $\textbf{k}_{\perp i},$
the convolution with the light-front wavefunctions and the integration over the
relative transverse momentum
then projects out the $L_z=0$ component of the
light-front wavefunctions with minimal $n$ --  the hadron distribution
amplitudes
$\phi_H(x,Q).$

In this Letter 
we shall calculate the cross section for 
$e^+e^-\to H \mathcal{G}_{J=0,2}$ using pQCD factorization. 
The amplitude at leading twist can be expressed as a
factorized product of the perturbative hard-scattering amplitude
$T_H(\gamma^* \to Q \bar Q g g)$ convoluted with
the nonperturbative distribution amplitudes for the heavy quarkonium and
glueball states.
We shall find that  $\gamma^* \to J/\psi \mathcal{G}_0$ production
dominates over that of $J/\psi \mathcal{G}_2$, and show how the angular
distribution of the final state can be used to determine the angular momentum
$J$ and projection $J_z$ of the glueball.  We shall show that
only $J_z = \pm 2$ tensor states are produced by the pQCD
mechanism at leading twist.
A bound on the normalization of the distribution amplitude for the glueball state
can be extracted from a  resonance search by CUSB in
$\Upsilon\to \gamma X$~\cite{CUSB-gamma}.
We shall show that the
rate for $e^+e^-\to J/\psi \mathcal{G}_0$ production could be comparable to the
corresponding NRQCD prediction for $e^+e^-\to J/\psi\eta_c$ without exceeding
the CUSB bound from radiative $\Upsilon$ decay.

The distribution amplitude $\phi_H(x,Q)$ required for the formation of the $H$
in a hard process is directly related to the NRQCD matrix element
for the leptonic decay rate of $H$. Its $x$ dependence is
peaked at $x \sim 1/2.$  The key quantity which determines the
normalization of the $\gamma^* \to H \mathcal{G}_J$ processes is then
the distribution amplitude $\phi_{J}(x,Q)$ of the $\mathcal{G}_J$.  The
pQCD factorization picture provides a direct relation among the
various glueball production processes, as they all involve the
same process-independent $\phi_{J}(x,Q)$.  The $\phi_{J}(x,Q)$
can be determined phenomenologically by
fitting to the measured production rate of a glueball candidate.
In leading-twist approximation the spin structure of the two-gluon system in
hard-scattering amplitude becomes that of a massless spin-$J=0,2$ 
state. 
Therefore the field-theoretic definition of the 
$\phi_J(x,Q)$ in light-cone gauge reduces to~\cite{BGL}
\begin{eqnarray}
\phi_{J}(x,Q)&=&
\frac{F^J_{\alpha\beta}}{\sqrt{2(N_c^2-1)}}
\int\frac{d^2\textbf{k}_\perp dz^-d^2\textbf{z}_\perp}{(2\pi)^3k^+x(1-x)}
\nonumber\\&&\times
e^{-i(xk^+z^- - \textbf{k}_\perp\cdot \textbf{z}_\perp)}
\nonumber\\&&\times
\langle \mathcal{G}_{J}| \textrm{T}~G_a^{+\alpha}(0^+,z^-,\textbf{z}_\perp)
                       G_a^{+\beta }(0)|0\rangle,
\label{phi2}
\end{eqnarray}
where $x$ and $\textbf{k}_\perp$ are the 
light-cone momentum fraction and transverse momentum of a gluon
inside the $\mathcal{G}_{J}$ with  momentum 
$k=(k^+=n\cdot k,k^-=\bar{n}\cdot k, \textbf{0}_\perp)$ and mass 
$M_{\mathcal{G}_J}$.
The $S-$wave component is projected out by integrating over $\textbf{k}_\perp$.
The light-like vectors $n$ and $\bar{n}$ satisfy
$n^2=\bar{n}^2=0$ and $n\cdot\bar{n}=2$.
The tensor $F^J_{\alpha\beta}$ projects the massless spin-$J$ components;
they are defined by
$F^0_{\alpha\beta}=[-g_{\alpha\beta}+\frac{1}{2}(n_\alpha \bar{n}_\beta
+n_\beta \bar{n}_\alpha) ]/\sqrt{2}$ and
$F^2_{\alpha\beta}$
is the massless spin-2 polarization tensor $\epsilon_{\alpha\beta}$.
The glueball distribution amplitude can also be defined from the
two-gluon
light-front wavefunctions 
$\psi_{\mathcal{G}_J}(x,\textbf{k}_\perp,\lambda_i)$ with gluon
spin projection $\lambda_i=S^z_i = \pm 1,$ integrated over transverse
momentum in light-cone gauge $A^+=0.$

The relative rates for the production
of heavy scalar glueballs with higher radial number $N$ are determined by
the normalization of the corresponding glueball distribution amplitudes.
In effect, the integral of the distribution amplitude over $x$ is
the relativistic generalization of the Schr\"odinger wavefunction at the
origin.
Thus the
distribution amplitudes for the
$0^{++}$ glueballs tend to scale inversely with their mean radius 
$\langle r_N \rangle$.
According to bag models~\cite{Kuti}, 
$\langle r_N \rangle\sim 0.6 $ fm, independent of $N$,
suggesting equal rates for the heavier glueballs.
On the other hand, the virial theorem extended to the light-front formalism
suggests that mean transverse momentum and  $1/\langle r_N \rangle$ 
increase monotonically with glueball mass.  If this is the case,
then the production rate in the 
$\gamma^* \to H \mathcal{G}_J$ 
will tend
to increase for heavier glueball states, assuming that the  annihilation
energy $\sqrt s$  poses no phase-space restriction.
Lattice gauge theory and light-front Hamiltonian methods should eventually
determine the glueball distribution amplitudes, thus providing
consistency checks on the production mechanisms   considered here.

As noted above, the amplitude for
$\gamma^* \to H(p) \mathcal{G}_J(k)$ can be 
computed as the convolution of 
$T_H(\gamma^*\to Q \bar Q g g)$
with $\phi_J(x,Q)$ weighted by the NRQCD matrix element.
In leading twist $k^-$ is neglected and, thus the glueball momentum 
is approximated by $k=k^+\bar{n}/2$ in $T_H(\gamma^*\to Q \bar Q g g)$.
The resulting effective vertex $\mathcal{A}_J^{\mu}$ is~\cite{BGL}
\begin{eqnarray}
\mathcal{A}_0^{\mu}
&=&
-\frac{8ig_s^2ee_Qm^2_Q \sqrt{N_c^2-1} }{N_c k\cdot n p\cdot \bar{n}}
\left(\epsilon_H^\mu -\frac{\bar{n}^\mu k\cdot n \epsilon_H \cdot \bar{n}}
                           {2 p\cdot \bar{n}}\right)
\nonumber\\&&\times
\sqrt{ \frac{\langle O_1\rangle_{H}}{m_Q^3} }
\;I_0,
\label{A0-effective}
\end{eqnarray}
\begin{eqnarray}
\mathcal{A}_2^{\mu}
&=&
-\frac{4ig_s^2ee_Qm^2_Q \sqrt{2(N_c^2-1)} }{3N_c k\cdot n p\cdot \bar{n}}
\;
\epsilon_2^{\mu\nu}(\lambda_2) \,\epsilon^H_\nu
\nonumber\\&&\times
\sqrt{ \frac{\langle Q_1^1\rangle_{H}}{m_Q^7}}
\;I_2,
\label{A2-effective}
\end{eqnarray}
where $\mu$ and $\epsilon_H$
are the vector indices for the $\gamma^*$ and polarization vector
for the $H$, respectively. 
The mass and fractional charge of the 
heavy quark $Q$ are expressed as $m_Q$ and $e_Q$.
Here $\langle O_1 \rangle_{H}$ and
$\langle Q_1^1\rangle_{H}$  are 
the vacuum-saturated analogs of
NRQCD matrix elements $\langle O_1(^3S_1)\rangle_{H}$
and $\langle Q_1^1(^3S_1)\rangle_{H}$ for annihilation
decays defined in Refs.~\cite{BBL} and \cite{Bodwin:2002hg}, respectively.
To leading order in the heavy-quark velocity $v_Q$ 
in the quarkonium rest frame, the $\langle O_1(^3S_1)\rangle_{H}$  
is related to the radial wavefunction at the origin $R(0)$
in the color-singlet model~\cite{singlet-model}
and the decay constant $f_H$,
which is defined by $\langle 0|J^\mu_{\textrm{e.m.}}|H\rangle%
=2M_H e_Q f_H\epsilon^\mu_H$, as
$\langle O_1\rangle_H=\frac{N_c}{2\pi}|R(0)|^2 =2M_H f^2_{H}$.
The nonperturbative factors for $\mathcal{G}_J$ are written as
$I_0=\int_0^1 dx \phi_{0}(x,Q)$ and  
$I_2=\int_0^1 dx \phi_{2}(x,Q)/[x(1-x)]$.
In leading twist 
the valence gluons are collinear
and therefore the only allowed polarization states for 
$\mathcal{G}_2$ are $\lambda_2=\pm 2$.  For $\mathcal{G}_2$ production 
the longitudinal polarization is prohibited by Bose symmetry.
This is true for any production process 
for $\mathcal{G}_2$, for which  pQCD factorization is valid. 
The amplitude (\ref{A2-effective}) for $\mathcal{G}_2$ is proportional to
the factor $\langle Q_1^1\rangle_{H}/m_Q^7$ which is suppressed to
$\langle O_1 \rangle_{H}/m_Q^3$ by 
$v_Q^4$.
Therefore, in the remainder of this
Letter we  only consider $\mathcal{G}_0$; 
the analysis for $\mathcal{G}_2$ 
can be found in our forthcoming publication~\cite{BGL}.
Using the vertex (\ref{A0-effective}),
we obtain the width $\Gamma_0$ for radiative $\Upsilon$ decay into 
$\mathcal{G}_0$ as 
\begin{eqnarray}
\Gamma_0
&=&
\frac{16\pi^2\alpha_s^2 \alpha e_b^2(N_c^2-1)\Phi^\gamma_0}
     {3N_c^2 m_b}
\frac{\langle O_1\rangle_\Upsilon}{m_b^3}
|I_0|^2,
\label{Gamma0}
\end{eqnarray}
where $\Phi^\gamma_0=1-M_{\mathcal{G}_0}^2/M_\Upsilon^2$.
In Ref.~\cite{He:2002hr}, the decay rate for the process
$\Upsilon\to \gamma f_0$ has been calculated treating $f_0$ as a glueball
candidate. The $\Gamma_0$ agrees with Eq.~(5) of Ref.~\cite{He:2002hr},
after including a missing factor $2/3$
and neglecting $M_{\mathcal{G}_0}$~\cite{Ma}.

Our result for the differential cross section for $e^+e^-\to J/\psi
\mathcal{G}_0$ normalized to
$\sigma_{\mu^+\mu^-}=4\pi\alpha^2/(3s)$ is
\begin{eqnarray}
\frac{dR_{J/\psi\mathcal{G}_0}}{d\cos\theta^*}
&=&
\frac{3\pi^2\alpha_s^2 e_c^2(N_c^2-1) r^2 \Phi^{ee}_0 }
     {N_c^2 \left(1-\frac{r^2}{4}\right)^2}
\frac{\langle O_1\rangle_{J/\psi}}{m_c^3}
\nonumber\\
&&\times
\frac{|I_0|^2}{s}
\left[ \sin^2\theta^*+\frac{r^2}{4}(1+\cos^2\theta^*) \right],
\label{dR0}
\end{eqnarray}
where 
$\theta^*$ is the scattering angle in the center-of-mass frame,
$r=4m_c/\sqrt{s}$, 
and the phase-space factor $\Phi^{ee}_0$ is defined by
\begin{eqnarray}
\Phi^{ee}_0=\frac{1}{s}
\sqrt{
\left[s-(M_{J/\psi}+M_{\mathcal{G}_0})^2\right]
\left[s-(M_{J/\psi}-M_{\mathcal{G}_0})^2\right]
          }.
\end{eqnarray}
The angular factors in the expression (\ref{dR0}) can be understood physically.
If the hadron pair is produced at $\theta^*=0$; i.e., aligned with the
lepton beams, then only final states with $J_z= \pm 1$ can contribute, because
the $e^+$ and $e^-$ annihilate
with opposite chirality.  Thus in the case of scalar glueballs, 
the $J/\psi$ with helicity $\pm 1$ is produced with a $1+ \cos^2 \theta^*$
distribution.  
If the $J/\psi$ is longitudinally polarized, the cross section
must vanish in the forward direction, and thus it has a  $\sin^2\theta^*$
distribution.

The rate integrated over angle is
\begin{eqnarray}
R_{J/\psi \mathcal{G}_0}
=
\frac{32\pi^2\alpha_s^2 e_c^2r^2 (1+\frac{r^2}{2})\Phi^{ee}_0}
     {9\left(1-\frac{r^2}{4}\right)^2}
\frac{\langle O_1\rangle_{J/\psi}}{m_c^3}
\frac{|I_0|^2}{s}.
\label{R0}
\end{eqnarray}
The size of the cross section can be estimated using the asymptotic form of
the ratio $\mathcal{R}=R_{J/\psi\mathcal{G}_0}/R_{J/\psi\eta_c}$
\begin{eqnarray}
\mathcal{R} \simeq \frac{9}{4}
\left(\frac{\alpha_s^{\mathcal{G}_0}}{\alpha_s^{\eta_c}}\right)^2
\frac{1+\frac{r^2}{2}}{r^2(1-r^2)(1-\frac{r^2}{4})^2}
\frac{m_c|I_0|^2}{\langle O_1\rangle_{\eta_c}}, \label{R0e}
\end{eqnarray}
where we neglected QED contributions to $R_{J/\psi\eta_c}$ given
in Ref.~\cite{Braaten:2002fi}. In the ratio $\mathcal{R}$ the
phase-space factor $\Phi^{ee}_0$ cancels the $\sqrt{1-r^2}$ for
$e^+e^-\to J/\psi\eta_c$. The 
$\alpha_s$'s
for the
two processes are written distinctively because they have
different effective scales.  However, the main uncertainties from
the choice of running coupling scale and scheme largely
cancel in the ratio $\mathcal{R}.$  Here
$\frac{m_c|I_0|^2}{\langle O_1\rangle_{\eta_c}}$ represents the
ratio of the square of the wavefunction at the origin of the
glueball compared to that of the $\eta_c.$

We next investigate whether some portion of the anomalously large signal for
$J/\psi+\eta_c$,
$\chi_{c0}$,  and $\eta_c(2S)$
observed by the Belle Collaboration could  actually be coming from the process 
$e^+e^-\to J/\psi\mathcal{G}_0$.  We   calculate the cross section assuming
glueball mass $M_{\mathcal{G}_0}$ 
the same as those for $\eta_c$, $\chi_{c0}$, and $\eta_c(2S)$.
In order to predict the production cross section $\sigma_{J/\psi\mathcal{G}_0}$,
we need to know the nonperturbative factors 
$\langle O_1\rangle_{J/\psi}$ and $I_0$.
The $\langle O_1\rangle_{J/\psi}$ is determined through 
the leptonic decay rate of $J/\psi$.  
As the glueball distribution amplitude is process independent, we can
extract an upper bound  to $I_0$ from the CUSB data for the resonance search 
from $\Upsilon\to\gamma X$.  We follow the method used in 
Ref.~\cite{Berger:2002bz}.  The branching fraction 
Br$[\gamma \mathcal{G}_0]$ for the process 
$\Upsilon\to\gamma \mathcal{G}_0$ is obtained by
\begin{eqnarray}
\textrm{Br}[\gamma \mathcal{G}_0]
=
\frac{\Gamma_0}
     {\Gamma[e^+e^-]_{\textrm{NRQCD}}}\times
        \textrm{Br}[e^+e^-]_{\textrm{exp.}},
\label{br}
\end{eqnarray}
where $\textrm{Br}[e^+e^-]_{\textrm{exp.}}=2.38$\% and
$\Gamma[e^+e^-]_{\textrm{NRQCD}}=%
2\pi e_b^2\alpha^2\langle O_1\rangle_\Upsilon/(3m_b^2)$.
In the ratio (\ref{br}) $\langle O_1\rangle_\Upsilon$ dependence
cancels.
The branching fraction must be less than allowed by the CUSB excluded region.
In order to extract the bound, we note that 
the mass resolution of the CUSB data is $20\,$MeV. 
If the  decay width $\Gamma[\mathcal{G}_0]$ of the $\mathcal{G}_0$
is larger than the resolution, one must rescale the  boundary of the 
excluded region by the factor $\Gamma[\mathcal{G}_0]/20\,$MeV.
The decay width $\Gamma[\mathcal{G}_0]$ cannot be 
computed using
perturbation theory because factorization is not valid for this
nonperturbative quantity.  However, if Belle's $J/\psi \eta_c$ signal
also contains $J/\psi \mathcal{G}_0$, $\Gamma[\mathcal{G}_0]$ must be
less than $110\,$MeV, which is the full width at half maximum of the $\eta_c$
peak in the Belle fit to the $J/\psi$ momentum distribution.
The first row in
Table~\ref{tab:I0} gives the upper limits to $|I_0|^2$
for 
$m_b=4.73\,$GeV and 
$M_{\mathcal{G}_0}=2.98$, $3.42$, and $3.65\,$GeV
corresponding to $M_{\mathcal{G}_0}=M_{\eta_c}$, $M_{\chi_{c0}}$, and
$M_{\eta_c(2S)}$, respectively.
Values for $|I_0|^2$ above the bound are excluded by $90$\% confidence level.
We choose $\alpha_s(\mu^2)=0.26$ 
using the modified minimal subtraction ($\overline{\textrm{MS}}$) scheme
and the scale 
$\mu^2 = e^{-5/3}\langle |\textbf{k}|^2\rangle$~\cite{Brodsky:1997dh} where
$\langle |\textbf{k}|^2\rangle$
is the mean 3-momentum squared for a single gluon.

\begin{table}[t]
\caption{\label{tab:I0}%
Upper limits to
the nonperturbative constant $|I_0|^2$,
cross section $\sigma_{J/\psi \mathcal{G}_0}$,
and the ratio
$\sigma_{J/\psi \mathcal{G}_0}/\sigma_{J/\psi h}$ at $\sqrt{s}=10.6$~GeV,
assuming $M_{\mathcal{G}_0}=M_h$, 
where $h=$ $\eta_c$, $\chi_{c0}$, and $\eta_c(2S)$.
The limits are determined by
the $\Upsilon\to\gamma X$ search of the CUSB
Collaboration~\cite{CUSB-gamma}.
}
\begin{ruledtabular}
\begin{tabular}{l|ccc}
$M_{\mathcal{G}_0}=M_h$ & $h=\eta_c$ & $\chi_{c0}$ & $\eta_c(2S)$
\\
\hline $|I_0|^2_{\textrm{max}}$ ($10^{-3}$~GeV$^2$)
& 5.2 & 5.8 & 6.2
\\
\hline $\sigma^{\textrm{max}}_{J/\psi \mathcal{G}_0}$
& 1.4 fb  & 1.5 fb  & 1.6 fb
\\
\hline $\sigma^{\textrm{max}}_{J/\psi \mathcal{G}_0}/%
        \sigma_{J/\psi h }$
& 0.63    & 0.72     & 1.9
\end{tabular}
\end{ruledtabular}
\end{table}

Now we are ready to find upper limits to 
$\sigma_{J/\psi\mathcal{G}_0}$ at the $B$-factories.
Substituting $|I_0|^2_{\textrm{max}}$ to Eq.~(\ref{R0}),
we get the cross sections 
$\sigma_{J/\psi \mathcal{G}_0}=R_{J/\psi\mathcal{G}_0}\,\sigma_{\mu^+\mu^-}$ 
in the second row in Table~\ref{tab:I0}.  
In order to make our prediction consistent with 
the previous analyses on exclusive charmonium-pair production,
we use the same input parameters given in 
Refs.~\cite{Braaten:2002fi,Bodwin:2002fk,Bodwin:2002kk}:
$\langle O_1\rangle_{J/\psi}=0.335\,$GeV$^3$,
$m_c=1.40\,$GeV, and $M_{J/\psi}=3.10\,$GeV.  The strong coupling constant
is chosen to be 
$\alpha_s=0.260$, $0.264$, and $0.265$ for 
$M_{\mathcal{G}_0}=M_{\eta_c}$, $M_{\chi_{c0}}$, and $M_{\eta_c(2S)}$,
respectively,
applying the same method used for the radiative $\Upsilon$ decay.
The ratio to the cross sections for exclusive charmonium-pair productions
are given in the third row in Table~\ref{tab:I0}.

The cross sections for $J/\psi+\eta_c$, $\chi_{c0}$, and $\eta_c(2S)$
recently measured by the Belle Collaboration are
not well understood within NRQCD.  Based on the assumption that
the measured signals at Belle include the $J/\psi+\mathcal{G}_0$ signal
within the mass region corresponding to $\eta_c$, $\chi_{c0}$, and $\eta_c(2S)$
we get the cross section for $J/\psi \mathcal{G}_0$.
We thus find that the upper limit to the cross section $\sigma_{J/\psi
\mathcal{G}_0}$ is comparable to the NRQCD prediction of the cross sections for
$e^+e^-\to J/\psi+h$  for $h=\eta_c$ and 
$\chi_{c0}$, and larger by factor 2 to that for $h=\eta_{c}(2S),$ suggesting the
possibility that a significant fraction of the anomalously large cross section
measured by Belle may be due to  glueballs in association with $J/\psi$
production.  In fact, there is a possibility of a resonance signal in the Belle
data for $e^+ e^- \to J/\psi X$ at the missing mass ${\cal M}_X \sim 1.7 $ GeV.
A resonance search in the radiative
$\Upsilon(nS)$ decay by the CLEO Collaboration and an independent study
by the BaBar Collaboration on charmonium-pair production
in $e^+e^-$ annihilation will provide stringent tests of this scenario.


\begin{acknowledgments}
We thank Geoffrey Bodwin, Eric Braaten, and Carl Carlson for valuable
discussions and comments.
The work of S.J.B. is supported by 
the U.~S.~Department of Energy under 
contract numbers DE-AC03-76SF00515 and DE-AC05-84ER40150.
The work of A.S.G. is supported in part by
the U.~S.~National Science Foundation under
contract number PHY-0140192, and by 
the U.~S.~Department of Energy under
contract number
DF-FC02-94ER40818.
The research of J.L. in the HEP Division
at Argonne National Laboratory is supported by
the U.~S.~Department of Energy, Division of High Energy Physics, under
contract W-31-109-ENG-38.
\end{acknowledgments}


\end{document}